\documentclass[
 reprint,
 amsmath,amssymb,
 aps]{revtex4-1}
\usepackage{comment}%
\usepackage{graphicx}% Include figure files
\usepackage{dcolumn}
\usepackage{bm}
\usepackage{multirow} % For fancy tables

\usepackage{xcolor}
\usepackage{cancel}

\def\be{\begin{equation}}
\def\ee{\end{equation}}
\def\bea{\begin{eqnarray}}
\def\eea{\end{eqnarray}}

\def\n{\ensuremath{\bm{n} }}
\def\nperp{\ensuremath{\bm{n}^* }}
\def\r{\ensuremath{\bm{r} }}
\def\d{\mathrm {d}}
\def\nl{\hfil\break} %new line
\usepackage[UKenglish]{babel}

\draft % marks overfull lines with a black rule on the right

\begin{document}

%\preprint{}

\title{Shape programming lines of concentrated Gaussian curvature } %Title of paper



\author{}
%\thanks{}
\affiliation{}

\author{D.\ Duffy}
\author{L.\ Cmok}
\altaffiliation[Also at ]{Jo\v{z}ef Stefan Institute, Department of Complex Matter, Jamova 39, SI-1000 Ljubljana, Slovenia}
\author{J.\ S.\ Biggins}
\affiliation{Department of Engineering, University of Cambridge, Trumpington St., Cambridge CB2 1PZ, U.K.}
\author{A.\ Krishna}
\altaffiliation[Also at ]{Center for Systems Biology Dresden (CSBD), Dresden 01307, Germany}
\author{C.\ D.\ Modes}
\altaffiliation[Also at ]{Center for Systems Biology Dresden (CSBD), Dresden 01307, Germany}
\altaffiliation[Also at ]{Cluster of Excellence, Physics of Life, TU Dresden, Dresden 01307, Germany}
\affiliation{Max Planck Institute for Molecular Cell Biology and Genetics (MPI-CBG), Dresden 01307, Germany.}
\author{M.\ K.\ Abdelrahman}
\altaffiliation[Also at ]{Department of Materials Science and Engineering, Texas A\&M University, College Station, TX 77843, USA}
\author{M.\ Javed}
\altaffiliation[Also at ]{Department of Biomedical Engineering, Texas A\&M University, College Station, TX 77843, USA}
\author{T.\ H.\ Ware}
\altaffiliation[Also at ]{Department of Materials Science and Engineering, Texas A\&M University, College Station, TX 77843, USA}
\altaffiliation[Also at ]{Department of Biomedical Engineering, Texas A\&M University, College Station, TX 77843, USA}

\affiliation{Department of Bioengineering, University of Texas at Dallas, Richardson, TX 75080, USA }
\author{F.\ Feng}
\author{M.\ Warner}
\email{mw141@cam.ac.uk}
\affiliation{Department of Physics, University of Cambridge, 19 JJ Thomson Ave., Cambridge CB3 0HE, U.K.}

%\affiliation{
% This line break forced% with \\}%
%\affiliation{
% Third institution, the second for Charlie Author}%


\date{\today}

\begin{abstract}
Liquid crystal elastomers (LCEs) can undergo large reversible contractions along their nematic director upon heating or illumination. A spatially patterned director within a flat LCE sheet thus encodes a pattern of contraction on heating, which can morph the sheet into a curved shell, akin to how a pattern of growth sculpts a developing organism. Here we consider, theoretically, numerically and experimentally, patterns constructed from regions of radial and circular director, which, in isolation, would form cones and anticones. The resultant surfaces contain curved ridges with sharp $V$-shaped cross-sections, associated with the boundaries between regions in the patterns. Such ridges may be created in positively and negatively curved variants and, since they bear Gauss curvature (quantified here via the Gauss-Bonnet theorem), they cannot be flattened without energetically prohibitive stretch. Our experiments and numerics highlight that, although such ridges cannot be flattened isometrically, they can deform isometrically by trading the (singular) curvature of the $V$ angle against the (finite) curvature of the ridge line. Furthermore, in finite thickness sheets, the sharp ridges are inevitably non-isometrically blunted to relieve bend, resulting in a modest smearing out of the encoded singular Gauss curvature.  We close by discussing the use of such features as actuating linear features, such as probes, tongues and grippers. We speculate on similarities between these patterns of shape change and those found during the morphogenesis of several biological systems.

\end{abstract}


\maketitle %\maketitle must follow title, authors, abstract and \pacs

\section{Introduction}\label{sect:intro}
%\label{}
Active shape-changing materials \cite{mcevoy2015materials, mirvakili2018artificial} are enticing engineering analogues for contracting muscles and growing tissues, and open up new vistas of soft bio-inspired machines \cite{cianchetti2014soft, gao20164d}. These materials typically undergo a simple but large local shape change in response to a stimulus: for example a gel will dilate isotropically on swelling, and a liquid crystal elastomer (LCE) will contract uniaxially along its nematic director on heating or illumination. However, just as differential growth and differential muscular contraction can generate exquisite and complex shape changes in biology \cite{thompson1942growth}, one can engineer complex global shape changes into LCE \cite{ware2015voxelated} and gel \cite{klein2007shaping} monoliths by spatially patterning the degree or direction of their local shape change \cite{warner2020topographic}. There is now a pressing need to elucidate the  geometric and mechanical principles that govern such active shape changes, both to gain insights into biology \cite{cancer, brainpnas}, and to facilitate soft machines.

Here we focus on shape-programming of flat nematic elastomer \cite{warner2007liquid} sheets, with thickness $t$ and encoded with a (unit) planar nematic director $\n(x,y)$. Upon activation by heating or illumination, the nematic molecular alignment is disrupted, and the sheet  contracts by a factor of $\lambda\sim 0.7$ along $\n$. This contraction is accompanied by an elongation of $\lambda^{-\nu}$ in both lateral directions, where the opto-thermal Poisson ratio is strictly $\nu=1/2$ in our incompressible elastomers, but may rise as high as $\nu=2$ in nematic photo-glasses. Activation of the flat sheet changes the in-surface distances between points, and hence the sheet's metric. More precisely, an unactivated infinitesimal length element  $\mathrm{d}\mathbf{l}=(\mathrm{d}x,\mathrm{d}y)$ has length $\mathrm{d}l^2=\mathrm{d}\mathbf{l}\cdot I \cdot\mathrm{d}\mathbf{l}$, which upon activation becomes
\be
\mathrm{d}\mathbf{l}_A^2=\mathrm{d}\mathbf{l}\cdot( \lambda^2 \n\n+\lambda^{-2 \nu} \nperp \nperp)\cdot\mathrm{d}\mathbf{l} \equiv \mathrm{d}\mathbf{l}\cdot\bar{a}\cdot\mathrm{d}\mathbf{l}\label{eq:stretches}
\ee
where $\nperp$ is the (in-plane) orthogonal dual of $\n$. Thus, the encoded director $\n(x,y)$  encodes a new metric for the sheet $\bar{a}$, which defines the intrinsic geometry of the actuated surface if infinitely thin (bend plays no role compared with stretch) and embeddable. 

The mechanics of such actuating sheets is dominated by two fundamental principles. Firstly, isometric bending occurs at a much lower energy scale ($\propto t^3$) than stretching ($\propto t$). Secondly, according to Gauss's \emph{theorema egregium} \cite{gauss1828disquisitiones, o2014elementary}, the Gauss curvature (GC) of a  surface (calculated as the product of its principle curvatures, $K=1/(R_1 R_2)$) is an intrinsic property metric, and cannot be modified by isometric bending. Strong actuation is thus achieved by encoding an $\bar{a}$ that bears GC, since blocking actuation will then incur energetically expensive stretch.

\begin{figure}[t]
	\centering
	\includegraphics[trim={0cm 0cm 0.4cm 0cm}, clip,width=\columnwidth]{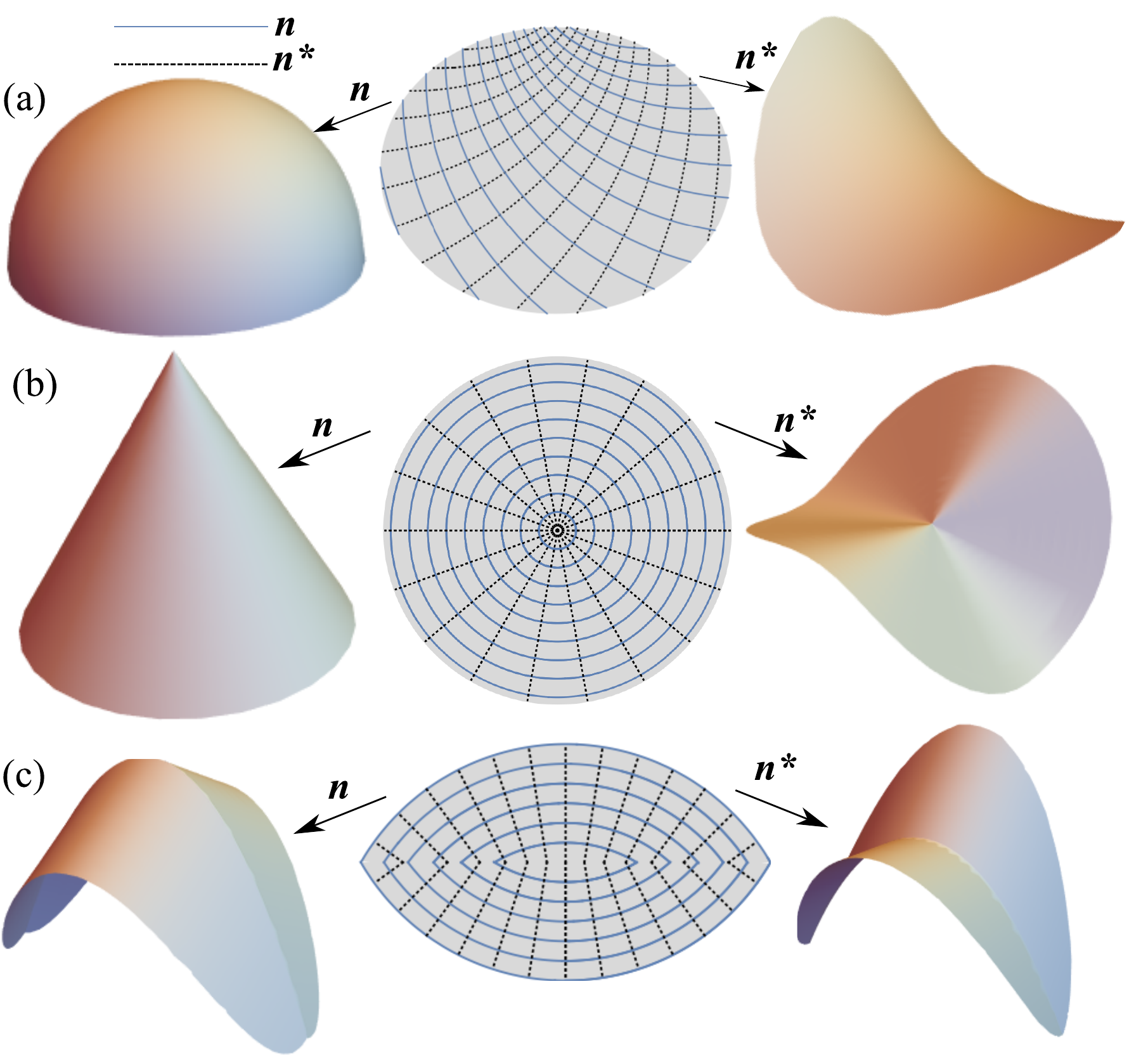}
	\caption{Director patterns $\n(\r)$ (with their orthogonal duals $\nperp(\r)$) are the programs that direct metric-driven morphing. (a) Distributed Gaussian curvature, without singular points or lines, from taking cuts from a suitable spiral pattern -- here spherical caps and saddles. (b) Circular $\n$ (radial $\nperp$) gives cones on contraction along $\n$, and anticones if the contraction is instead along $\nperp$, both of which are Gaussian flat except at points -- the tip (anti-tip). (c) Director pattern yielding lines of positive (negative for the dual) Gaussian curvature, which are the subject of this investigation.}
	\label{fig:programming}
\end{figure}

Previous work on actuating sheets has focused on two types of Gauss curvature. Geometrically, the most elementary case is distributed GC, which is locally finite, and leads to smooth surfaces on actuation. For example, LCE and gel disks programmed to yield uniform positive Gauss curvature morph into spherical caps on activation \cite{klein2007shaping,aharoni2014geometry,mostajeran2015curvature,kowalski2018curvature}. In the LCE case, one can also consider a sheet prepared with the orthogonal dual director pattern, $\n \to \nperp$. Such sheets have the the same metric form as the original (eqn.\ \ref{eq:stretches}) but with $\lambda$ and $\lambda^{-\nu}$ exchanged, indicating the exchange of the contractile and elongational directions. This transformation reverses finite Gauss curvature $K_A \to -K_A$ \cite{mostajeran2015curvature}. For example, as seen in Fig.~\ref{fig:programming}(a),  the dual of a uniform positive Gauss curvature sheet encodes uniform negative  curvature, and sculpts the sheet into a  saddle. Similarly, if one cools rather than heats and LCE sheet, one gets elongation along $\n$ and contraction along $\nperp$, again producing a saddle.

At the other geometric extreme, one can instead program metrics with points of singular Gauss curvature, leading to surfaces with sharp points \cite{modes2010disclinations, modes2011gaussian, de2012engineering,defective_nematogenesis}, but which are otherwise Gaussian-flat. Paradigmatically, if a nematic disk is encoded with concentric director rings (a simple $m = 1$ defect when confined to 2-D) then, on activation, every circumference reduces by a factor of $\lambda$, and every (in-material) radius rises by $\lambda^{-\nu}$, morphing the disk into a cone with semi-angle $\sin{\phi}=\lambda/\lambda^{-\nu}$, as seen in Fig.~\ref{fig:programming}(b). The resultant cone is Gauss flat everywhere, except at the sharp tip which bears a positive integrated curvature $\Omega\equiv\int K_A \mathrm{d}A_A=2\pi (1-\lambda/\lambda^{-\nu})$. Though confined to a single point, this finite integrated curvature imbues the cone with great strength:  White {\it et al} \cite{guin2018layered-rev} have demonstrated that such LCE cones can lift thousands of times their own weight as they rise. In this case of Fig.~\ref{fig:programming}(b), the orthogonal dual pattern is an LCE disk encoded with instead a radial director, which loses radius and gains circumference on actuation, and thus buckles into a ruff shape. Again, since this pattern is the orthogonal dual of concentric circles, and hence equivalent but with $\lambda$ and $\lambda^{-\nu}$ exchanged, the surface is an \emph{anticone} that is Gauss flat except with a point of negative integrated curvature $\Omega\equiv\int K_A \mathrm{d}A_A=2\pi (1-\lambda^{-\nu}/\lambda)$ at the origin.

In this paper, we consider the final, intermediate possibility: metrics that encode lines of singular Gauss curvature, leading to surfaces that contain curved ridges with a sharp $V$-shaped cross-sections as sketched in Fig.~\ref{fig:programming}(c). Though superficially similar to the curved folds deployed in origami, these ridges are intrinsic to the metric, and cannot be removed isometrically. Given the spectacular performance of point sources of GC as lifters, we are motivated to study lines of GC to give line lifters instead. Moreover, such intrinsically sharp lines are a common sight in biology. It is also the case in morphogenesis that the structures grow from an initially flat tissue, thus exhibiting a shape change similar to the activating LCE sheets we focus on here. We speculate further at the end of the paper.

Here we design elementary nematic director patterns that morph flat LCE sheets into surfaces containing such ridges, quantify their encoded Gauss curvature, and then explore, experimentally and numerically, the true shapes of the resultant surfaces. We find that ridges come in positively and negatively curved variants (again produced by dual director fields, Fig.~\ref{fig:programming}(c)) which are distinguished by whether the (finite) curvature of the ridge line is in the same or opposite sense as the (divergent) transverse curvature of the ridge's $V$.  Our numerics and experiments highlight two key features of such intrinsic ridges: first, to some degree, changes in the $V$ angle may be accommodated isometrically at the cost of additional curvature along the ridge line, and, secondly, finite thickness effects mean the sharp tip of the $V$ is blunted (non-isometrically) to relieve the singular bend at the cost of some stretch. Finally, 
% we expand on some mechanisms that could play a role in the growth of the biological ridges mentioned previously, and
we speculate about the mechanical properties of intrinsically curved ridges. In particular, our speculations suggest LCE ridges may be useful to induce stiffness in actuating linear features such as probes, grabbers and perhaps biological elements.

\section{Patterns encoding Gaussian curvature concentrated on lines}\label{sec:patterns}
\begin{figure}[!ht]
	\centering
	\includegraphics[trim={0cm 1.0cm 0.9cm 0cm}, clip,width=\columnwidth]{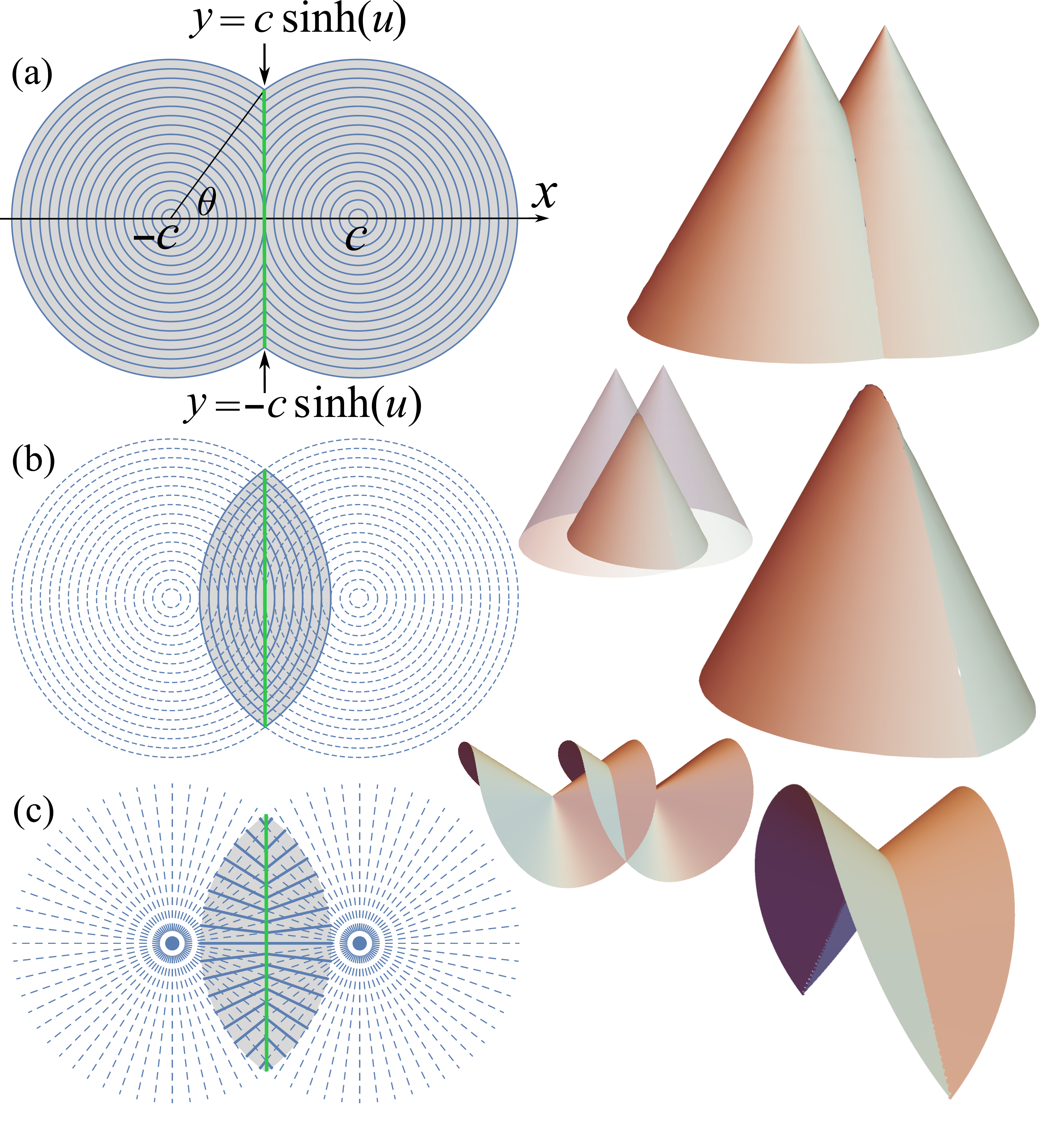}
	\caption{(a) Two circular director patterns, centred at $(\pm c, 0)$, intersecting along their bisector, actuate to a pair of fused cones. The (rank-1 connected) interface between the patterns, shown in green, becomes the negatively curved ridge between the cones on actuation. (b) On actuation, the complement of pattern (a) gives the complement of the fused-cones --- that is, the surface formed from the overlapped portions of the cone flanks discarded in (a). The resulting pitta-bread like surface has a positively curved sharp ridge. (c) The orthogonal dual of (b) (an anticone-complement pattern)  generates a surface which loosely corresponds to the overlapped region from a pair of anticones, and correspondingly bears a negatively curved ridge.}
	\label{fig:twocones}
\end{figure}

We start by designing nematic director patterns that will morph a flat sheet into a surface that is Gauss flat except along a sharply folded curved line. We take as our elementary building block a concentric circle director pattern, which generates a cone on activation which is Gauss flat everywhere except at the tip. If we fuse two circle patterns together along the bisector of their centers (which are placed at $x=\pm c$), the LCE sheet will then morph into a fused pair of cones on actuation, as shown in Fig.~\ref{fig:twocones}(a) \cite{fengbigginswarner}, and the (hyperbolic) folded ridge between the cones is indeed a line bearing finite integrated Gauss curvature. Intuitively, the ridge appears to bear negative GC, since the hyperbolic line curves in the opposite sense to the infinite curvature of the ridge's $V$. Indeed, asymptotically far from the two tips the fused conical surface looks like a simple single cone, so the entire hyperbolic fold must bear total integrated curvature that is the negative of one tip (distributed in some manner along its length) confirming that this is a negatively curved line. 

 We note that the line of concentrated curvature is associated with a line of discontinuity in the director field. This is perhaps expected, since the \emph{theorema egregium} provides an expression for the Gauss curvature in terms of derivatives of the metric, and hence derivatives of the director \cite{aharoni2014geometry,mostajeran2015curvature, defective_nematogenesis}. Along such a boundary, the meeting angle of the director with the boundary must be the same from each side to ensure that a boundary element actuates to the same length under the actuation of both sides: a metric condition known as rank-1 connectedness \cite{MWSPIE:12}. The bisecting boundary considered here clearly satisfies this condition, as, for the same pair of patterns, do a whole family of hyperbolic boundaries that generate pairs of cones with different heights \cite{fengbigginswarner}. These boundaries are easily described in an elliptical coordinate system characterised by the parameter $u$ along hyperbolae, a degenerate case being shown in Fig.~\ref{fig:twocones}(a). 

However, this cone-fusion surface also has Gauss curvature at the cone tips. The ridge may be isolated by cutting out a strip of pattern  that does not extend to the tips. However, a more satisfactory alternative is to consider the \emph{cone-complement} pattern (Fig.~\ref{fig:twocones}(b)) formed by flipping which circle pattern is deployed in each half of the sheet. The actuated surface is now the complement of the fused cones, consisting of the previously discarded pair of cone-flanks lurking beneath the seam line. These flanks form a pitta bread-like activated surface (Fig.~\ref{fig:twocones}(b)) with an intrinsically positively curved ridge in an otherwise Gauss flat landscape that may now be extended to infinity without encountering other sources of GC.

 As in previous cases, we expect the orthogonal dual pattern to form a ridge with opposite curvature, as shown in Fig.\ \ref{fig:programming}(c). Taking the orthogonal dual converts the  concentric director circles into director radii, and the conical building blocks into anticones. The resultant surfaces may thus be expected to resemble the intersection and complement of a pair of overlapping anticones, (Fig.~\ref{fig:twocones}(c)), which indeed leads to a positively curved ridge in the intersection pattern, and a negatively curved ridge for the complement. However, we emphasize that the deformation underlying even a single anticone is rather complicated, including an azimuthal component of displacement, so this construction serves only as a guide rather than as an exact isometry.
 
 \section{Experimental and numerical realisation of lines of positive and negative curvature}\label{sec:exp}

\begin{figure*}[!ht]
	\centering
	\includegraphics[trim={0cm 0cm 1.15cm 0cm}, clip,width=1.99\columnwidth]{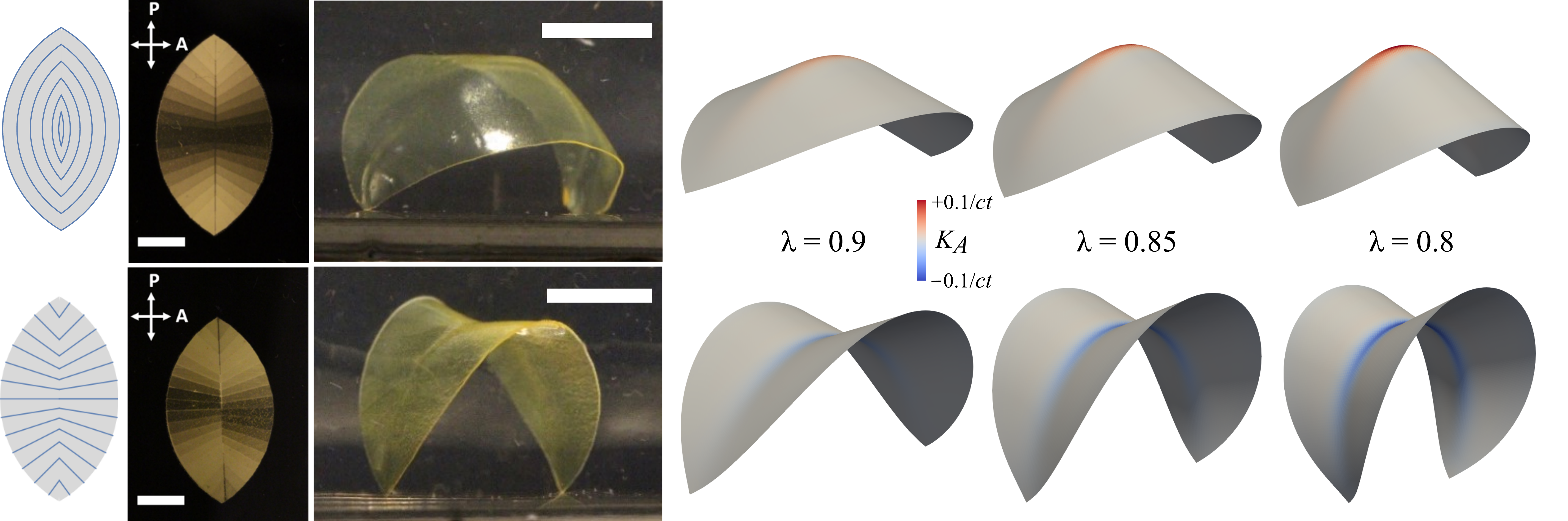}
	\caption{Left to right: Director patterns, images of the prepared LCE films with the film placed between crossed polarizers, experimental photographs taken at 150 $^{\circ}$C in silicone oil, and simulations at a range of $\lambda$ values, for cone complement (top) and anticone complement (bottom) cutouts. All scale bars are 5 mm, and simulations are coloured by GC.}
	\label{fig:experimpics}
\end{figure*}

In Section~\ref{sec:patterns} we identified cone and anticone complement patterns as prototypical examples of flat sheets that morph into surfaces containing singular lines of positive/negative GC. As a first step towards understanding the  mechanical properties of such line-like actuators, we realize these patterns experimentally and observe their emergent forms. Our experiments deployed 50$\mu$m thick LCE sheets  formed by photo-patterning the director pattern into a nematic fluid, then UV cross-linking to form an elastomer \cite{ware2015voxelated,dualresponsive}, (details in appendix \ref{sec:exp_details}). Upon releasing the film and heating to 150 $^{\circ}$C a contraction $\lambda =0.9 - 0.85$ was obtained  along the director, forming the actuated surfaces shown in Fig. \ref{fig:experimpics}. To further clarify the forms of the surfaces, we also conducted matching numerical calculations using a bespoke active-shell code MorphoShell~\cite{defective_nematogenesis} (details in appendix \ref{sec:num_details}) that resolves both the bending and stretching of the actuating LCE sheet.

As seen in Fig.\ \ref{fig:experimpics}, the cone and anticone complement patterns indeed yield qualitatively different surfaces, bearing curved folds with positive and negative GC respectively. Furthermore, the experiment and numerical surfaces are in good agreement about the overall shape of the actuated surfaces. Interestingly, although the surfaces are qualitatively similar to those anticipated theoretically in Fig.\ \ref{fig:twocones}, they differ in two important ways. Firstly, the curved fold is not actually singularly sharp, but rather somewhat blunted. Secondly, the perimeter of the cone complement surface is not planar like the analytic pitta-bread cone-flank construction in Fig.\ \ref{fig:twocones}, indicating that the opening angle of the fold is somewhat different. In what follows, we will argue that the first effect is a non-isometric relaxation to relieve the singular bend energy of a sharp fold, while the second effect is an isometric relaxation achieved by simultaneously increasing the curvature of the fold line to preserve the (singular) GC. We anticipate that these two forms of relaxation will dominate the mechanics of any actuator based on singular lines of GC. 

 \section{Curvature distribution along lines} \label{sect:GCdist}
Before exploring these two forms of shape relaxation in cone/anticone complement patterns, we first characterize the distribution of GC along their ridges. At first sight this may seem a hopeless task as the GC on the line is singular. However, just as a cone has a finite integrated curvature $\Omega\equiv\int K_A \mathrm{d}A_A$ arising from the singular curvature at its tip, so these lines have a finite integrated curvature distributed along their length. To characterize the curvature of such a line we thus seek the total integrated curvature given by the entire line, $\Omega$, which is a dimensionless quantity, and also the concentration of $\Omega$ per unit length along the line, which will be a quantity with dimensions of $1/\mathrm{length}$.

A simple way to calculate integrated curvature is provided by the Gauss-Bonnet theorem \cite{o2014elementary}, which states that the integrated curvature of any surface is
\[
\Omega\equiv \int K_A \mathrm{d}A_A= 2 \pi \chi - \oint k_g \mathrm{d}s,
\]
where $\chi$ is the topological Euler characteristic of the surface, and $k_g$ is the geodesic curvature of the surface's boundary. The geodesic curvature of a line on a surface is calculated as the projection of the line's curvature vector into the surface, and, like Gauss curvature, is an intrinsic geometric property that is unchanged by isometric bending deformations. 

We first consider the circular patterns actuating into the fused twin cones of Fig.~\ref{fig:twocones}(a). The valley separating the actuated cones develops negative Gaussian curvature, and a simple argument for the very large cone case \cite{MWSPIE:12} can quantify this GC contribution:  An isolated cone has integrated GC  at its tip of $2\pi(1 - \lambda^{1+\nu})$.   A twin cone, very large compared with the separation of its peaks, and seen from the outside, appears as a single cone. Its skirt tends to being circular, so, in Gauss-Bonnet, $\oint k_g \mathrm{d}s$ around the skirts must give the same value as for a single cone, indicating that the total GC enclosed is the single tip value. One thus concludes that an integrated GC of  $- 2\pi(1 -\lambda^{1+\nu})$ has been accrued in addition to that of the two peaks, and that this negative GC, that cancels one peak's worth of GC, is concentrated in the curved-fold of the landscape.

To probe the distribution of the curvature along the length of the fold, one can explicitly apply Gauss-Bonnet to a pair of finite fused cones whose extent is defined by the hyperbolic parameter $u$ as seen in Fig.~\ref{fig:twocones}(a). By computing the integrated GC in the crease for any value of $u$, and hence in any length segment of crease, we may then deduce the curvature GC concentration along the crease itself. For the finite case, the application of Gauss-Bonnet is somewhat involved, as one must take careful account of the finite contributions to the geodesic curvature integral at the point where the crease touches the boundary (see Appendix \ref{sec_appendix_GB}). However, we have already seen that the integrated GC associated with the tips is partially cancelled by the total GC associated with the crease. Applying Gauss-Bonnet to find the total integrated GC inside the outer boundary of the fused cones, and taking away the $2\times 2\pi(1-\lambda^{1+\nu})$ contribution from the tips  one ultimately obtains for the crease contribution  where 
\be  \label{eq:creaseGC}
\begin{split}
\Omega =\int & K_A \mathrm{d}A_A =  4\left( \vphantom{\frac12} \lambda^{1+\nu}\cos^{-1}(\mathrm{sech}(u)) \right. \\
& \left.-\cos^{-1}\left(\frac{\lambda^{1+\nu}}{\sqrt{\lambda^{2(1+\nu)}+  \sinh^2(u)}}\right)   \right)   \end{split}
\ee
 The $ \Omega$ value yielded via Eqn.~(\ref{eq:creaseGC}) indeed takes the negative of one tip's GC in the large limit $u \gg 1$, as previously anticipated.
 
 For the cone-complement pattern in Fig.~\ref{fig:twocones}(b), an exactly corresponding calculation via GB reveals that the ridge bears exactly equal and opposite GC. This result follows naturally from the pitta-bread form of the surface, since the ridge line follows the same hyperbola as the cone fusion ridge, while the opening angle of the $V$ is exactly reversed. Therefore the finite principal curvature of the surface is unchanged while the singular one is exactly reversed, overall resulting in a negation of the Gauss curvature $K=1/(R_1 R_2)$.

% see MW geodesic_subtraction.nb - in folder /Feng
Having calculated the integrated curvature for any value of $u$, we  now characterize its distribution along the length of the crease --- that is, the integrated curvature $\Omega$ per unit length along the positively curved cone complement crease. This quantity corresponds to integrating the GC only transversely to the crease to yield a lineal density along the crease of integrated GC, $\rho$ say, a quantity of dimension 1/length. When shape programming active materials, one has two measures of this integrated GC density, according to whether one measures per unit length of the seam in the reference state, that is $\rho_y$, or per unit length of the crease in the actuated (target) state, $\rho_{s_A}$, where $s_A$ is arc length along the actuated crease. The latter captures the singular curvature of the line in the target state, and defines its target state geometry, while the former captures where the curvature is encoded in the reference state pattern. 

\begin{figure}[h]
	\centering
	\includegraphics[trim={0cm 0.0cm 0cm 0cm}, clip, width=\columnwidth ]{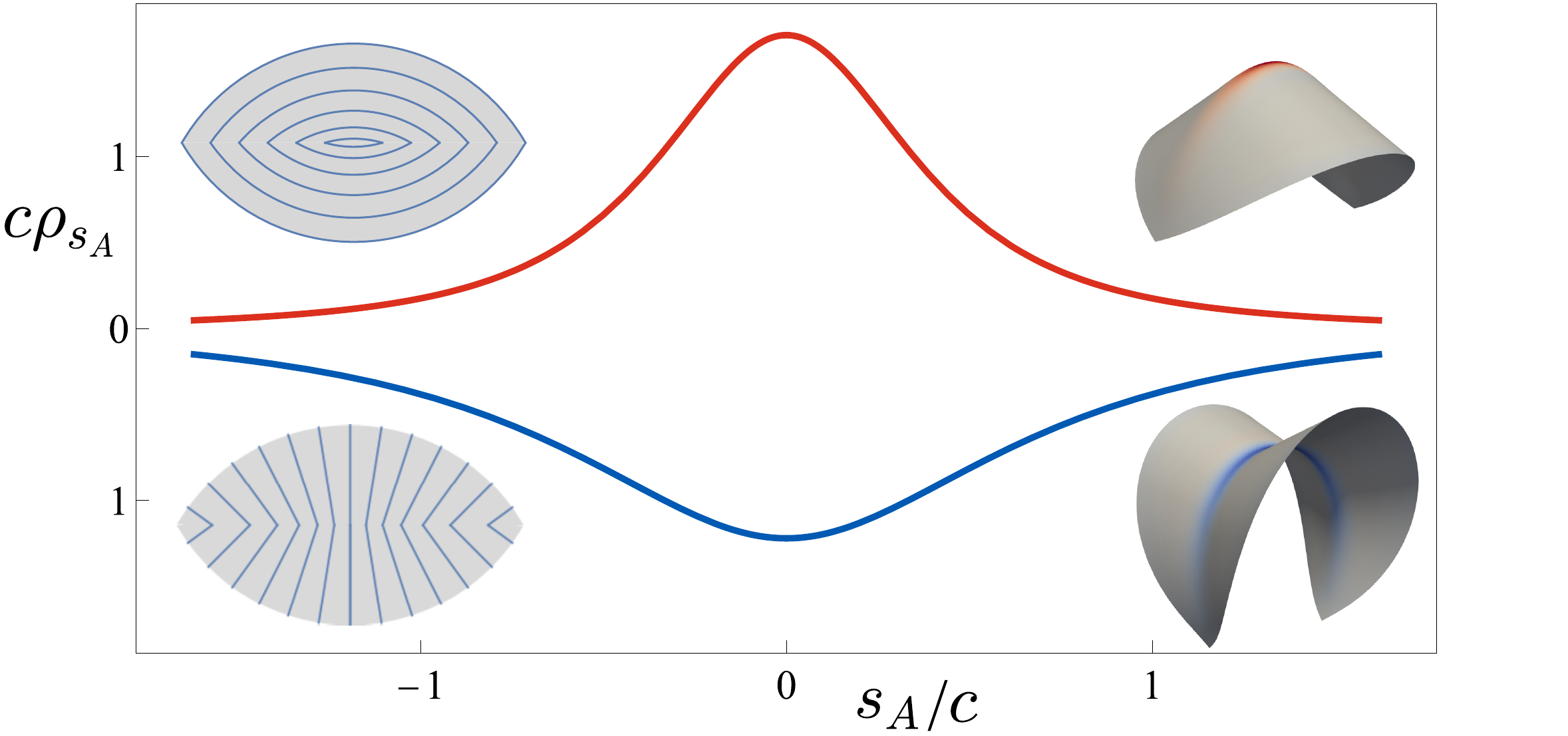}
	\caption{Lineal Gauss curvature density $\rho_{s_A}$ of cone complement (red curve) and anticone complement (blue curve) patterns, with parameters (given in Table~\ref{table:sims}) corresponding to the $\lambda=0.8$ cutouts in Fig.~\ref{fig:experimpics}, which are shown again here.}\label{fig:theory_fig}
\end{figure}

The integration leading to  $\Omega$ of eqn~(\ref{eq:creaseGC}) is for the full fused-cone ridge between $y=\pm c \sinh{u}$, so we may identify $\Omega_h=-\Omega/2$ as the integrated curvature in the cone-complement half-ridge extending from the the central apex to the edge, i.e.\ between the points corresponding to $y=0$ to $y=+c \sinh(u)$ in the reference domain. For the density of once-integrated curvature along the crease, we need to take the derivative of $\Omega_h$ with respect to the appropriate length in either the reference or target state. Since $\Omega_h$ is known as a function of the hyperbolic parameter $u$, this requires the appropriate Jacobeans $\d y/\d u = c \cosh(u)$ and $\d s_A/\d u = c \sqrt{ \lambda^{2}  + \lambda^{-2\nu} \sinh^2(u)}$. The respective measures of once-(transverse)-integrated GC densities along the cone-complement ridge are thus
\bea\label{eq:creaseGClocal-y}
\rho_y \!=\! \frac{\partial \Omega_h}{\partial y} \!=&\!  \frac{2\lambda^{1+\nu} }{c} (1- \lambda^{2(1+\nu)}) \frac{\mathrm{sech}^2(u)}{\left( \lambda^{2(1+\nu)} + \sinh^2(u) \right) }\\
\label{eq:creaseGClocal-sA}
\rho_{s_A}  \!=\!\frac{\partial \Omega_h}{\partial s_A} \!=& \frac{2\lambda^{1+2\nu} }{c}(1- \lambda^{2(1+\nu)}) \frac{  \mathrm{sech}(u)}{\left( \lambda^{2(1+\nu)} + \sinh^2(u) \right)^{3/2}}.
\eea
In Fig.\ \ref{fig:theory_fig}, we plot the ridge lineal curvature density $\rho_{s_A}$ parametrically against $s_A = - \textrm{i} c \lambda \textrm{EllipticE}(\textrm{i}u,1/ \lambda^{2(1+\nu)})$ for a cone complement pattern actuated with $\lambda=0.8$. We see that the GC is indeed positive, concentrated near the central apex of the ridge, and decays to zero at distances $\gg c$ from the center.

Since the anticone complement pattern is the orthogonal dual of the cone complement, we may trivially obtain its Gauss curvature distribution from the above results by exchanging $\lambda$ and $\lambda^{-\nu}$, and hence setting $\lambda^{1+\nu}\to \lambda^{-{1-\nu}}$ in eqn.\ \ref{eq:creaseGC} 
and eqn.\ \ref{eq:creaseGClocal-y}, and also $\lambda^{1+2 \nu} \to  \lambda^{-\nu-2}$ in eqn.\ \ref{eq:creaseGClocal-sA}. As anticipated in the introduction, this substitution changes the factor $(1- \lambda^{2(1+\nu)})>0$ in equations \ref{eq:creaseGClocal-y} and \ref{eq:creaseGClocal-sA} to $(1- 1/\lambda^{2(1+\nu)}) < 0$, reversing the sign of the integrated GC density. In Fig.\ \ref{fig:theory_fig}, we also plot the ridge lineal curvature density $\rho_{s_A}$ for a matching anticone complement pattern. Comparison of the plot for these two dual system's highlights this reversal, but also, curiously, clarifies that, although the sign of the GC has flipped, the magnitude also changes so the curvature is not perfectly reversed. Perfect reversal is always seen in LCE sheets programmed with smooth director fields (leading to smooth surfaces)\cite{mostajeran2015curvature}, but not at the sharp (anti)conical tips generated by $m=1$ defect patterns \cite{modes2011gaussian}. This discrepancy is indicative of a fundamental difference between finite and singular curvature in LCE shape programming \cite{defective_nematogenesis}, and highlights the singular nature of the curvature programmed along these lines.

\section{Isometric and non-isometric relaxation of singular ridges}
Our above analyses are based on exact isometries of the programmed metrics constructed from sections of cones and contain ideally sharp ridges that are surely geometric idealizations. One can imagine two ways in which a physical LCE sheet may relax from these idealized shapes: it may bend into a different surface that is still an isometry of the activated metric, or it may stretch into a new surface with a different metric. In any thin sheet, deviations from isometry are penalized by a stretching energy (per unit area) $\propto \mu t$, while  bending with curvature $1/R$ incurs a bending energy $\propto \mu t^3/R^2$. Consequently, thin sheets are normally expected to adopt an exact isometry of the activated metric, to minimize the dominant stretching energy, while the bending energy is relegated to a `tie breaker' between different available isometries. 

\begin{figure}[h]
	\centering
	\includegraphics[trim={0.7cm 0.5cm 0.0cm 3.4cm}, clip, width=\columnwidth ]{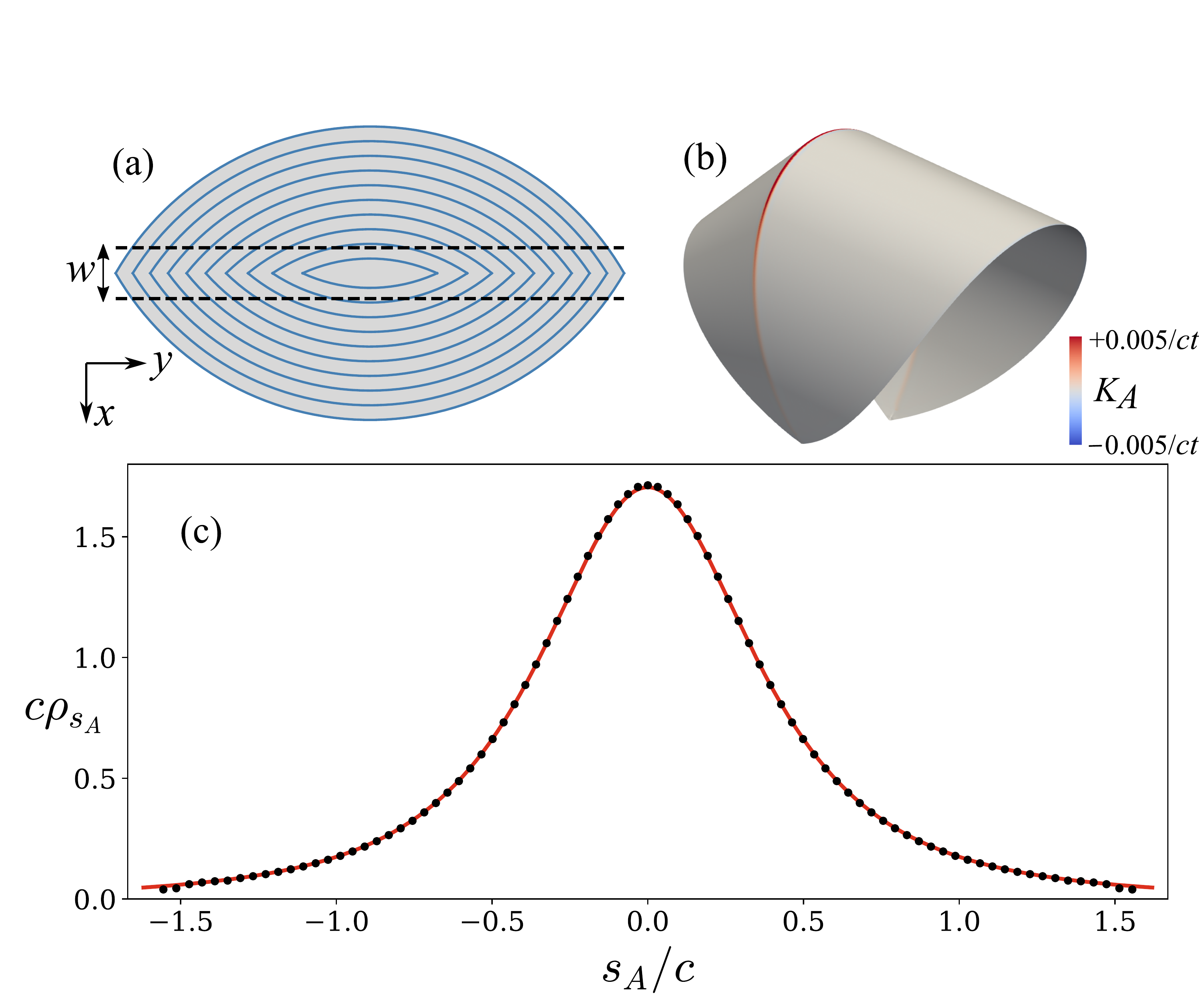}
	\caption{(a) A sketch of a cone complement director pattern. (b) The corresponding simulated cutout, with $\lambda = 0.8$, $\nu = 1/2$, and further parameters given in Table~\ref{table:sims}. (c) Plot of $\rho_{s_A}$ against $s_A$, with numerics (black dots) agreeing well with the theoretical (red) curve, eqn.~(\ref{eq:creaseGClocal-y}). The non-zero thickness $t$ in the simulation regularises the GC, so it is finite and non-isometrically `smeared' over a narrow region around the crease. The numerical $\rho_{s_A}$ was therefore calculated by integrating the GC in the transverse direction, over an interval (of width $w$ in the reference state) large enough to comfortably encompass the full non-isometric region.}\label{fig:simulation1}
\end{figure}

However, at a sharp ridge, the bending energy is divergent as the transverse curvature is infinite. Around the ridge, the bending energy may thus compete with the stretching energy, giving rise to non-isometric relaxations that blunt the ridge by smearing the singular transverse curvature over a small transverse lengthscale. This basic mechanics is familiar from the rounding of LCE cone tips \cite{modes2011gaussian} and results in the  large but finite curvature values seen along the ridges in Fig.\ \ref{fig:experimpics}. Since such blunting arises from a trade off between bend and stretch, the structures must sharpen as the sheet is made thinner and attain a truly sharp isometry in the limit $t \to 0$. In Fig. \ref{fig:simulation1} we present a further simulation on an identical cone-complement actuator but with thickness reduced by a factor of 50, which clearly demonstrates this sharpening.

It is natural then, to wonder whether this blunting results in a fundamental change to the GC distribution along the ridge. To investigate, we compute the (large but finite) GC distribution on our simulated surface via the angular deficits at the underlying mesh's vertices, and then numerically integrate along a short transverse distance, to get the total curvature per unit activated length along the ridge $\rho_{s_A}$. As seen in Fig. \ref{fig:simulation1}, this numerical calculation is in almost perfect agreement with the analytic result in eqn.\ \ref{eq:creaseGClocal-y}. In hindsight, this result is perhaps unsurprising. Gauss-Bonnet shows that the integrated GC within any patch of surface can be deduced from the geodesic curvature (and hence the metric) of the patch boundary. Thus if we apply GB to a large patch that entirely contains the ridge, the patch boundary lies in regular isometric material, so the GC within must be unchanged: blunting of the ridge can smear out the singular curvature over the blunted region, but cannot change the overall amount.

 However, the simulated and experimental cone-complement shapes still differ markedly from the ideal ``pitta-bread'' isometry: there is a marked discrepancy in both the fold angle and the (finite) ridge curvature, which lead to the boundary of the physical samples being far from planar. The excellent agreement between idealised and simulated $\rho_{s_A}$ indicates that the simulated shapes are essentially isometries of the programmed metric, but with blunted ridges. This large scale global discrepancy must therefore be via an isometry that redistributes bend without stretching. Additional evidence for this view is found in the low values of simulated stretching energies far from the ridge line (see Fig.~\ref{fig:energy_fig}), and from the fact that the thinner simulations in Fig. \ref{fig:simulation1} deviate \textit{further} from the analytic shape than their thicker counterparts in Fig.\ \ref{fig:experimpics}, despite being closer to perfect isometry.
 
\begin{figure}[h]
	\centering
	\includegraphics[trim={0.4cm 0.2cm 0.9cm 0.4cm},clip, width=\columnwidth ]{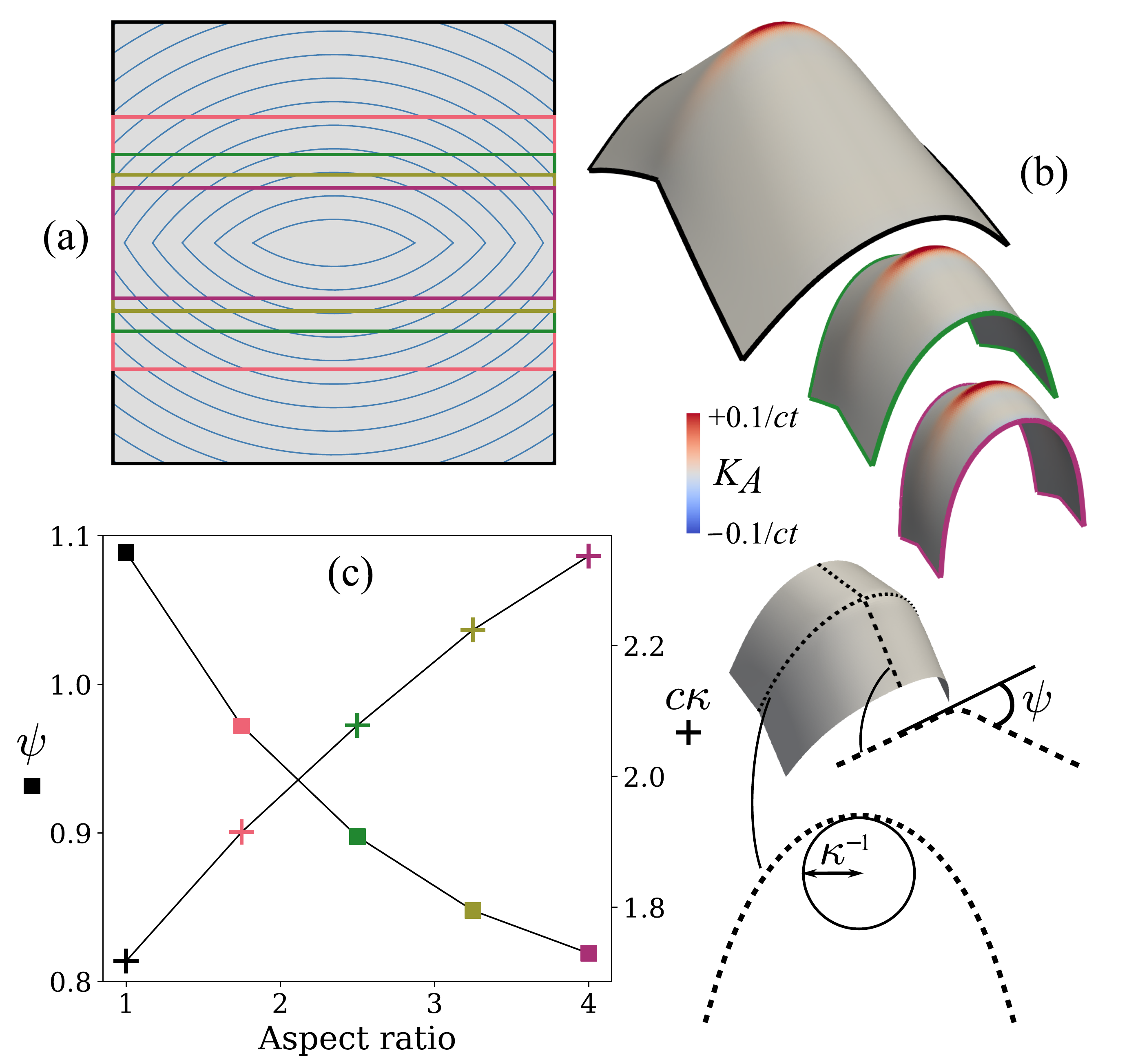}
	\caption{(a) A cone complement director pattern from which five rectangular reference domains of varying aspect ratio are `cut out'. (b) Simulations of three of the corresponding activated cutouts, coloured by Gauss curvature, with $\lambda=0.8$, $\nu=1/2$. (c) A plot of both transverse `fold' angle $\psi$ and apex curvature of the crease line $\kappa$, against aspect ratio. This shows that creases with smaller fold angles must curve \textit{more} in the longitudinal direction, to preserve $\rho_{s_A}$, the (once-integrated) GC prescribed by the encoded metric.}\label{fig:simulation2}
\end{figure}

 To understand this relaxation we recall that, although Gauss curvature $K=1/(R_1 R_2)$ is an intrinsic property that cannot be modified without stretch, its constituent orthogonal principal curvatures, are not: one may change $R_1$ and $R_2$ in isometric bending deformations provided their product remains the same. In the case of a surface with a curved sharp ridge, this suggests the $V$ angle of the ridge is not fixed but may change isometrically as long as there is a sympathetic change in the (finite) transverse curvature of the ridge line. These different states are all (blunted) isometries of the activated metric, and selection between them is based on the resultant elastic energy, which will involve bend and stretch contributions along the blunted ridge, but just bend in the isometric flanks. 

  We illustrate this trade off numerically in Fig.\ \ref{fig:simulation2} by simulating a series of cone-complement patterns containing the same length of ridge, but with different transverse extents. Our calculations reveal that the samples with the smallest width adopt less folded shapes with more ridge curvature, while the wider samples are more tightly folded and less curved. For sufficiently thick samples such as ours,  the energetic motivation for this difference is that curvature of the ridge line generates curving and bend energy throughout the flanks of the sheet, whereas the singular bend of the $V$ imposes energies only in a sharp vicinity of the ridge. Thus, in wider sheets it is preferable to pay for a sharper fold to mitigate bend in the flanks. While in a narrow strip, which contains just as much ridge but less flank, the cost of ridge curvature is correspondingly lower, so a less folded and more curved isometry is advantageous.

The mechanics of  wide variety of non-folded strips and ribbons have been considered in the literature, for instance the collection \cite{ribbons-collection}. These studies include strips imbued with geodesic curvature\cite{dias2012geometric}, and finite distributed Gauss curvature \cite{Efrati_2015}, but, to our knowledge, none consider ribbons with lines of concentrated Gauss curvature as considered here.  Our narrowest strip is  superficially reminiscent of a folded paper ribbon \cite{fuchs1999more, dias2012geometric, dias2014non}, generated by excising a narrow strip from around a curved origami fold. However, the mechanics of our strip is fundamentally different, as the central fold is intrinsic to the sheets metric and may not be removed isometrically, whereas a paper ribbon can always be unfolded. Consequently, in folded paper ribbons, a fold is only seen in elastic equilibrium because plastic deformation during folding endows the line with a local energetically preferred opening angle, modeled as an angular spring along the fold. In the narrow ribbon limit, this spring energy dominates a paper ribbon, and the preferred opening angle is obtained, whereas in wider ribbons the consequent bending of the ribbon flanks competes, leading to reduction in the degree of folding. In contrast, in our case it is the metric itself that encodes the crease, causing a global stretching energy in the flanks if the crease is flattened. This difference is highlighted by the opposite behavior of our ribbons, which becomes more folded as the width increases, as outlined above.

\section{Conclusions and mechanical/biological speculations}
\subsection{Conclusions}
We have explored the shape programming of initially flat LCE sheets that, on heating or illumination, form surfaces with Gauss curvature concentrated along lines. This contrasts with previous work that has focused on curvature concentrated at points, and finite curvature distributed in an areal manner. The surfaces that arise bear sharp curved ridges with a $V$ shaped cross-section that are intrinsic to the programmed metric, meaning they cannot be flattened without energetically prohibitive stretch. Such ridges come in positively and negatively curved variants, depending on the relative sign of the divergent curvature of the $V$ and the finite curvature of the ridge line.  Furthermore, although the GC at these ridges is singular, the integrated Gauss curvature $\Omega=\int K_A \mathrm{d}A_A$, and hence the integrated curvture per unit length along the ridge, are both finite intrinsic quantities that can be computed via the Gauss-Bonnet theorem. 

Here, we have focused on elementary examples of such ridges, programmed into the LCE by combining patches of $m=1$ defect nematic director patterns that individually morph into (anti)conical building blocks.  Our simplest director pattern --- the  \emph{cone-complement} --- encodes a positively curved ridge, while its orthogonal dual, an \emph{anticone-complement}, encodes a negatively curved ridge.

Simulation and experiment demonstrate that such singular ridges relax non-isometrically into a blunted form to relieve the divergent bend of sharp, idealised ridges. However, this highly localised diffusion of Gaussian curvature does not change the lineal distribution of curvature along the ridge, and the shapes attained by the sheets are essentially isometries of the programmed metric, albeit with a blunted apex. However, even such intrinsic ridges can undergo isometric deformations, in which the angle of the $V$ is traded against the curvature of the ridge line whilst maintaining the lineal Gauss curvature distribution. Indeed, both experimental and simulated LCE sheets exploit these isometries, and bend away from our simple analytic forms constructed from portions of cones. Interestingly, this global isometric relaxation trades the bend energy throughout the sheet and the stretch/bend energy associated with the blunted ridge, resulting in an overall shape that depends on both the LCEs thickness and its planar extent.

The patterning strategy where the liquid crystal director in solid sheets is programmed to give these responses in synthetic systems has close analogies in morphogenesis in biological systems where the patterning is either by anisotropic elongation and growth or by the localization and gradients of morphogens. We therefore conclude this paper with some short speculations about the formation of similar structures in biology, and the mechanical deployment of such structures as LCE actuators. Looking ahead, we also anticipate that the future holds many fruitful instances of cross-fertilization between these two analogous fields of study.

\subsection{Possible connections with morphogenesis.}
\begin{figure}[!ht]
	\centering
	\includegraphics[width=\columnwidth]{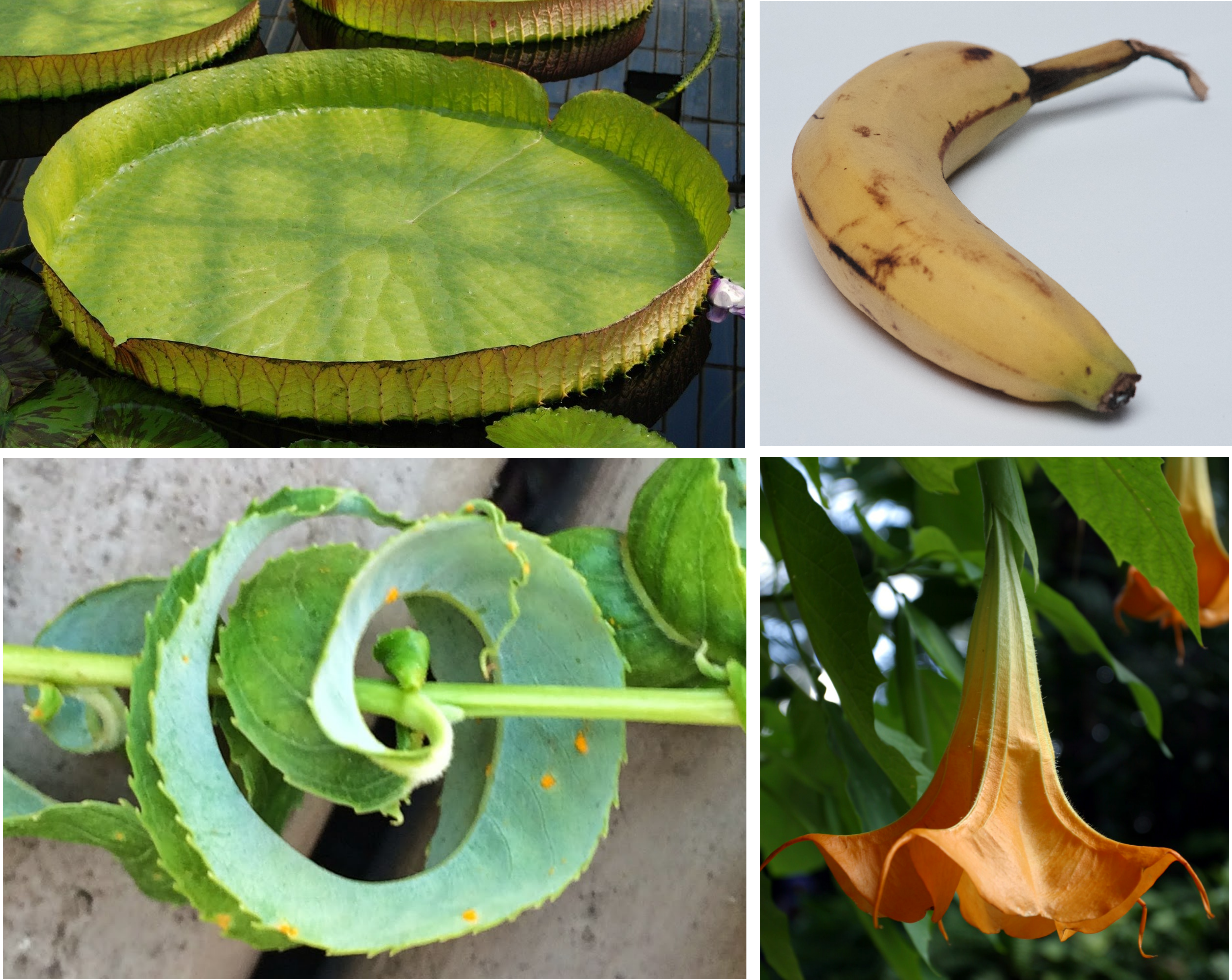}
	\caption{Lines of concentrated positive (+) and negative (-) GC in biology. Clockwise from top-left: A giant water lily (+), a banana (+ and -),  an Angel's Trumpet flower (-) ~\cite{angelstrumpet}, \textit{Salix babylonica} `Annularis' (-)~\cite{salixbabylonicaannularis}.}
	\label{fig:intro_bio}
\end{figure}

Fig.\ \ref{fig:intro_bio} illustrates that shells containing lines of concentrated Gauss curvature are commonly found across a huge range in biology.
For instance, a banana exhibits lines of GC of opposite signs in two different regions, similar to a torus which has regions of positive and negative GC. Lines of negative GC can be found in leaves like that of the decorative plant \textit{Salix babylonica} `Annularis' \cite{karp2011genetic}, while a line of positive GC is found around the circumference of a giant water lily pad.  In animals, margins of appendages like wings or fins exhibit a highly curved strip of tissue that joins the two flat sides. Examples include the pupal wing of \textit{Drosophila melanogaster} \cite{waddington1940genetic, de2017forces} and the pectoral fin of zebrafish \cite{grandel1998development}.

What are the parallels between the morphogenesis of these biological structures and the ``nematogenesis'' of our LCE sheets?
In general, changes to the 3D shapes of biological structures rely on (combinations of) three basic cellular effects:  shape change, division and rearrangement \cite{julicher2017emergence}. During development, these  simple local shape changes must be suitably spatially patterned to encode a desired complex global shape change, much like we have designed patterns of nematic contraction to produce our curved ridges. One may thus ask what patterns of shape change evolution deploys to create curved ridges, and how the appropriate pattern of growth is stimulated within the tissue.

The morphogenesis of plant tissues is particularly close to the shape programming of LCEs, as plant tissues are largely constructed from 2D sheets, and develop without cell rearrangement. Indeed, on occasions, patterns of cellular orientation within a leaf or petal encode patterns of elongation/contraction during growth that are strikingly similar to LCE director fields. For example, the iconic nectar spur of Darwin's Orchid, \emph{Angraecum sesquipedale}, forms from a flat disk of cells oriented in concentric circles, like a +1 nematic defect. During spur development, the cells elongate radially, causing the spur to protrude as a capped tube \cite{puzey2012evolution} exactly akin to the development of a cone from an LCE sheet. Similarly, in the Venus fly trap, elongation of aligned cells perpendicular to the curved midrib (driven by transient changes in turgor pressure) play an important role in the shape change of the leaf when it snaps \cite{hodick1989mechanism, forterre2005venus}. We speculate that  a similar mechanism could be at play to produce the striking curved folds found along the primary vein of some leaves, exemplified by the \textit{Salix babylonica} `Annularis' leaf in Fig.\ \ref{fig:intro_bio}. More precisely, in many flat leaves, cells or secondary veins are elongated at an angle to the primary vein. This angling could direct the deformation of the leaf in specific directions on changes in the turgor pressure or during development. The resulting  patterns of contraction/elongation would be strikingly reminiscent of the anticone complement director pattern, including being  naturally ``twinned" along the primary vein, and thus might similarly be expected to encode a GC onto the vein itself. 

\begin{figure}[]
	\centering
	\includegraphics[width=\columnwidth]{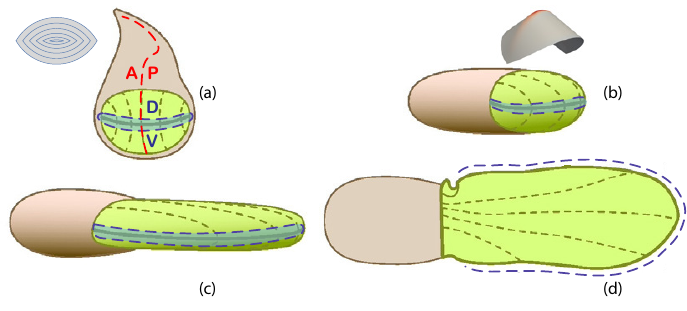}
	\caption{The \textit{Drosophila} wing imaginal disc consists of the pouch region (green) which grows into the wing blade of the adult fly. At the larval stage (a), the pouch is relatively flat and is divided into a dorsal (D) and a ventral (V) compartments by the DV boundary (blue) and the anterior (A) and posterior (P) compartments by the AP boundary (red). On eversion (b and c), the dorsal and ventral compartments face each other at the basal surface and flatten out while the DV boundary develops into a ridge with concentrated Gaussian curvature, reminiscent of the LCE sheets considered here and pictured above panel (b). The wing further undergoes growth through expansion, elongation and interaction with the hinge (brown) to grow into the adult wing (d). Figure adapted from Ref \cite{de2017forces}. M.C. Diaz de la Loza, B.J. Thompson, Forces shaping the Drosophila wing, Volume 144, Part A,
2017; licensed under a Creative Commons Attribution (CC BY) license.}
	\label{fig:wing}
\end{figure}

Moving beyond plants, a striking parallel of our cone-complement actuators is found in the development of a \textit{Drosophila} wing disc pouch into an adult wing. As seen in detail in Fig.\ \ref{fig:wing}, the wing pouch starts as a flat disk divided by a Dorsal-Ventral (DV) boundary which resembles the line of director discontinuity in the cone-complement. During development, the wing pouch morphs first into pitta-bread like shape (with the DV boundary indeed being the ridge) and then finally into a flat bi-layer structure in which the DV boundary has become completely closed and forms the wing margin. In this instance, we speculate that the patterned growth is templated by the localization and gradients of chemical morphogens within the tissue. Certainly several different chemical morphogens are known to pattern the imaginal disk, including concentration gradients of Hedgehog (Hh) and Decapentaplegic (Dpp) establish an Anterior-Posterior (AP) axis (left right in Fig.\ \ref{fig:wing}a), while Wingless (Wg) is localized on the Dorsal-Ventral boundary \cite{tabata2004morphogens}. Similarly, Bone Morphogenetic Protein (BMP) is expressed along the curved ridge of the pectoral fin of zebrafish \cite{mateus2020bmp}. Much remains to be learnt about the causal mechanisms linking such morphogen concentrations to tissue growth. However, at least in the \textit{Drosophila} wing, it is evident that the  complementary morphogen patterns combine to template the disk with the appropriate symmetry needed to encode a cone-complement-like pattern of growth.

\subsection{Mechanics, actuation and grabbers}
A strong motivation for studying the morphing of patterned LCE sheets, is to design soft actuators that can do useful mechanical work. For example, the GC encoded at the tip of LCE cones endows them with great strength, so they can lift large loads as they rise. We thus conclude our paper with some final speculations about the mechanics of surfaces with lines of GC, and their suitability for use as actuators.

\begin{figure}[h]
	\centering
	\includegraphics[width=0.9\columnwidth]{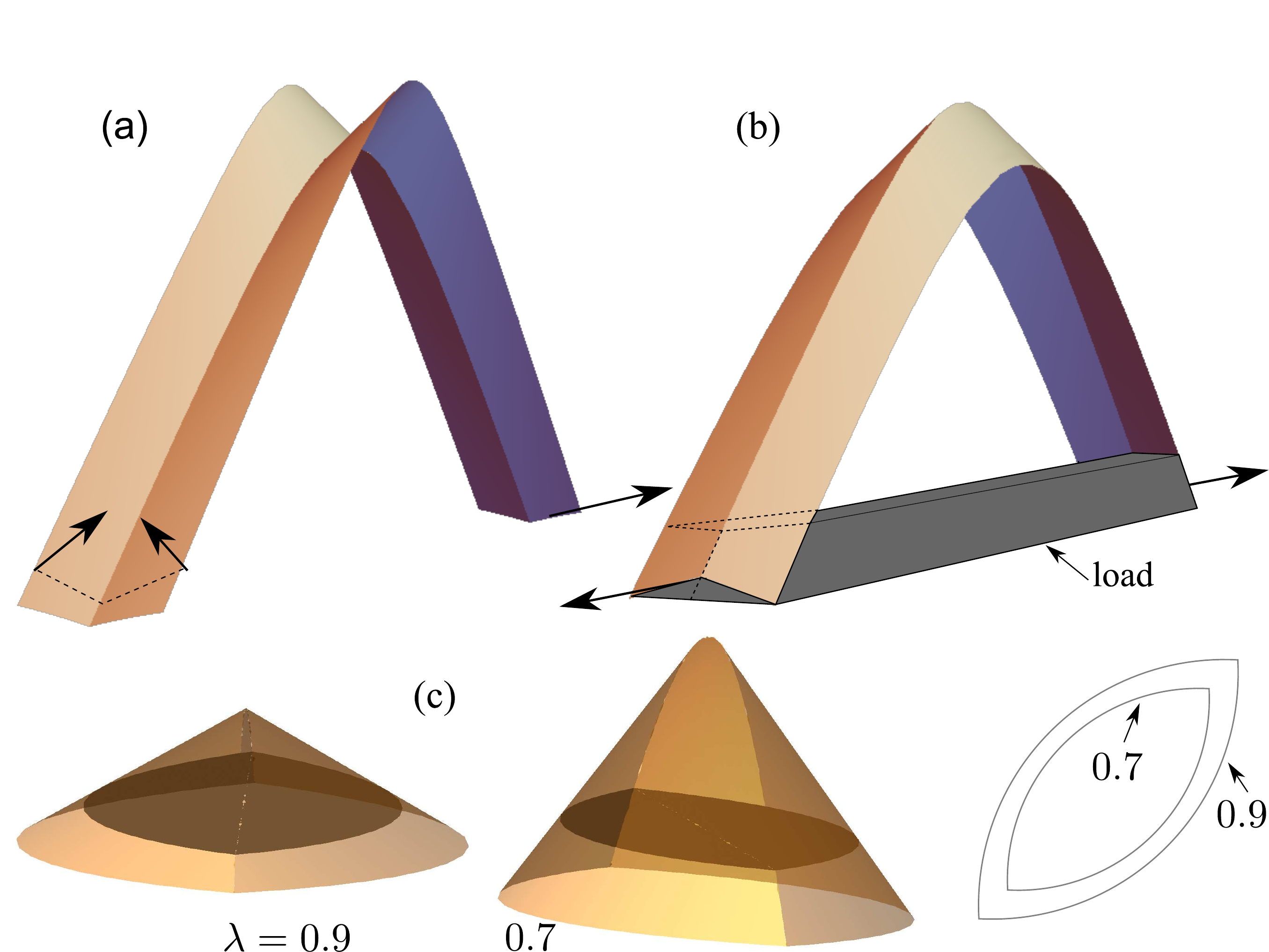}
	\caption{Curved beams with central lines of (a) negative GC, or  (b) positive GC. In (a) the movement of one end, indicated, is controlled by the adjustment of the V angle at the other end; closure of the V is shown. In (b) the arrows indicate a pair of outward directed forces either opening the ends apart, or here a grabber's load exerts blocking forces resisting beam closure as $\lambda$ is further reduced.  (c) A fuller version of the beam (b) acting as a Gaussian grabber to accentuate the stretch action in the flanks. The load fits at $\lambda = 0.9$, say (the enclosed shape just gripped). At $\lambda = 0.7$, middle, the load would have had to shrink (see final shapes) to fit the evolved cone complement, or flank stretches result.
	}
	\label{fig:mechanics}
\end{figure}

The cutouts of the fused or complementary director fields that we have considered here lead naturally to an exploration of the mechanics associated with beams containing longitudinal lines of concentrated GC. Curved shells are typically strong because their deformations involve changes of GC and hence stretch.  The field of active materials is dominated by bi-layers, which, like the original bimetallic strips, program shape change in the thickness direction to create bend. Such benders have been deployed as grabbers \cite{zeng2020associative}, lifters/pushers \cite{xiao2019biomimetic}, swimmers \cite{camacho2004fast}  and even robotic arms \cite{yamada2009photomobile}. Conventional beams are optimised for bend resistance by apportioning material away from the neutral plane, as in an I beam, but such structures have not been realized in active materials. Here, we suggest that the intrinsically folded ribbons introduced here could provide a bending-beam form of actuation, but endowed with the strength of shells and I-beams, as the arched ribbons cannot be flattened without stretch, causing a strong force  to be delivered to a blocking load. Figs.~\ref{fig:mechanics}(a) and (b) show such beams with positive and negative GC concentrated along their central seams, and would be suitable for use in all contexts where bending strips have previously been deployed.
% generated in D:\Users\mw141\Documents\Lab\science\Modes\origami\curved_fold\intersecting_cones.nb for .svg editing
% 

The recurrent theme in our crease analysis is that the GC is locally preserved and, since it is the product of the bend across the V and the bend along the crease, altering one bend influences the other. In a mechanical system, this offers potential for control and steering of the actuated structure.  The basic mechanism is sketched in Fig.~\ref{fig:mechanics}(a) where the nearest end has its V manipulated as shown by arrows -- closure towards the book geometry causes the crease to straighten and the other end to move in the direction arrowed, and vice versa. The actuated strip thus forms a steerable robotic arm, that could manipulate, sense or probe items in its vicinity.

We also see potential for lines of Gauss curvature in mechanical grippers. At a given $\lambda < 1$, the bent beam of Fig.~\ref{fig:mechanics}(b) could be partially straightened by the action of a force pair as shown. Again, the V has to close towards the book geometry. However, there is a hard limit to the crease straightening since, as we have analysed (see also Appendix~\ref{sec_appendix_GB}, especially eqn.~(\ref{eq:GB-applied})), the crease has geodesic curvature on each side (it is the difference between these curvatures which gives the GC). Gripping a load at this crease extension and further reducing $\lambda$ would lead to very strong stretch forces on the load.

Since an unobstructed LCE beam would bend strongly as $\lambda$ is reduced, the force pair shown in Fig.~\ref{fig:mechanics}(b) acts against the direction of closure under actuation, and the load therefore feels a corresponding pair of ``grabbing'' forces:
 Consider the grabber as in Fig.~\ref{fig:mechanics}(b) that has closed in on its load and engaged with it.  Because of the object, the crease cannot close its V where it is gripping the load. Also, increases of curvature along the crease are restricted by the load rigidity. The object cannot be accommodated by shape changes as GC needs to change with changing $\lambda$, and stretch must result.  Fig.~\ref{fig:mechanics}(c) shows a grabber with more complete flanks closing in at $\lambda = 0.9$, say, on a shaded object matching its cross section at a given distance down the flank. Were this load to deform with the grabber in the transition to $\lambda = 0.7$, say, then it would fit, as shown in the next panel. The difference between the load shape encountered at $\lambda = 0.9$ and what it would have to deform to in order to fit at $\lambda = 0.7$ is shown in the last panel. Clearly there are huge stretch effects in the flanks generated by the load mismatch if it is unyieldingly retaining its initial shape while the gripper is being heated to smaller $\lambda$.
 
 It is possible to have more complex patterning to weld together many such beams of the type of Fig.~\ref{fig:mechanics}(b). The GC creases radiate from higher order central defects. We are investigating more such new mechanics and actuation paradigms involving developing lines of GC. 
\vspace{.5cm}

	{\centering \bf Data Availablity 
	\par
	}

The data that support the findings of this study are available from the corresponding author upon reasonable request.

 \begin{acknowledgments}
F.F. and M.W. were supported by the EPSRC [grant number EP/P034616/1]. M.W. is grateful for support from the ELBE Visiting Faculty Program, Dresden. D.D. was supported by the EPSRC Centre for Doctoral Training in Computational Methods for Materials Science [grant no. EP/L015552/1]. J.S.B. was supported by a UKRI “future leaders fellowship” [grant number MR/S017186/1].\nl
This material is partially based upon work supported by the National Science Foundation under Grant DMR 2041671.
\end{acknowledgments}
 
 \appendix 
\section{Experimental Details}\label{sec:exp_details}
Films were synthesized by filling LCE precursors in a glass cell assembled from clean glass slides. Glass slides were cleaned for 5 minutes each in alconox-water solution, acetone, and isopropanol and exposed to oxygen plasma for 1 minute at a pressure of 100 mTorr with 50 W power. A solution of 1 wt\% brilliant yellow was prepared in dimethylformamide, filtered through a 0.45 $\mu$m pore size filter, and spin-coated on clean glass slides at 750 rpm with an acceleration of 1500 rpm/s for 10s and 1500 rpm with an acceleration of 1500 rpm/s for 30 s. A glass cell was constructed by adhering two dye-coated slides with a 50 $\mu$m spacer in between. Photoalignment of the azobenzene-based dye was induced by exposing the glass cell to broadband, linearly polarized light using a modified projector (Vivitek D912HD) with a patterning resolution of 30  $\mu$m. To obtain the desired pattern, selective regions of the cell were exposed to different polarization angles. The monomer solution was prepared by heating and vortexing a liquid crystal monomer, 1,4-bis-[4-(6-acryloyloxyhexyloxy)-benzoyloxy]-2-methylbenzene (RM82, Wilshire Chemicals) with n-butylamine (Sigma Aldrich) in the molar ratio 1.1:1 with 1.5 wt\% of the photoinitiator, Irgacure I-369 (BASF). The solution was filled into the glass cell by capillary action at 75$^{\circ}$C and stored at 65$^{\circ}$C for 12h to allow for oligomerisation. The sample was then cooled down to room temperature and exposed to UV light (OmniCure® LX400+, 250 mW/cm$^2$, 365 nm) for 2.5 minutes on each side for crosslinking. After crosslinking, the top glass slide was carefully removed so that the LCE films remain on the bottom glass slide. The patterned region in the film was laser cut from the surrounding polydomain regions with a CO$_2$ laser cutter (Universal Laser Systems ILS9.150D). Alignment was observed using polarized optical microscopy. To analyze actuation, samples were immersed in a silicone oil bath and heated to the desired temperature. Shape change was imaged using a Canon DSLR camera.

\section{Numerical Details}\label{sec:num_details}

Our numerical approach models a thin sheet of incompressible Neo-Hookean elastomer as a 2D surface, assigns an elastic energy to that surface, and minimises the energy to find equilibrium configurations. The energy penalises both stretching (deviations from the programmed metric $\bar a$, see eqn.~(\ref{eq:stretches})) and bending:
\begin{align} 
E &= \int \left( W_{\mathrm{stretch}} + W_{\mathrm{bend}} \right) \mathrm{d}A \label{eqn:energy} \\
  &=  \int \frac{\mu t}{2}\left({\mathrm{tr}}\left[
 a \, \bar a^{-1} \right]  + \frac{1}{\mathrm{det} [ a \, \bar{a}^{-1}]} - 3 \right)\mathrm{d}A \notag \\
 &+\int \frac{\mu t^3}{12\, \mathrm{det}[\bar{a}]} \left( \mathrm{tr}[(\bar{a}^{-1} \, b)^2] + (\mathrm{tr}[\bar{a}^{-1} \, b])^2 \right)\mathrm{d}A \notag
\end{align}
where $\mu$ is the shear modulus, and $t$ is the reference-state thickness. For a surface described by deformed-state position $\bm{r}(\bm{x})$ as a function of reference-state position $\bm{x}$, the metric $a = (\bm{\nabla} \bm{r})^\mathrm{T} \bm{\nabla} \bm{r}$. The `second fundamental form', which encodes the (extrinsic) bending, is given by $b = (\bm{\nabla} \bm{r})^\mathrm{T} \bm{\nabla} \bm{\hat{N}}$, where $\bm{\hat{N}}$ is the normal to the deformed surface. The stretching term is minimised at $a = \bar{a}$, but the bend energy is minimised by a planar surface for which $b=0$. Thus the bending energy in general leads to (localised) regions of non-isometry ($a \neq \bar{a}$), but also selects the isometry of lowest bending energy outside these regions.

The deforming 2D surface is approximated by a mesh of flat triangles, with each triangle assigned a single pair of matrices $a,\, b$, estimated from the local geometry. The metric $a$ is estimated from the unique linear deformation
that describes the current positions of the triangle’s three
nodes: a standard constant-strain finite-element approach. The second fundamental form $b$ involves second derivatives,
and hence is estimated from the unique quadratic deformation
consistent with the positions of six ‘patch’ nodes close to the triangle’s centroid. The energy~(\ref{eqn:energy}) is thus approximated by a discrete sum over triangles, which is minimised over the mesh node positions via damped Newtonian dynamics. For further details, and the Morphoshell code, see Ref.~\cite{defective_nematogenesis}.

We summarise various simulation parameters in Table~\ref{table:sims}, where we denote the seam length in the reference state by $L$, and aspect ratio by AR.

\begin{table}[h!]
\centering
\caption{Parameters used for the simulations appearing in the main text. All simulations had $\nu=1/2$, $L/c=3.46$. In Fig.~\ref{fig:simulation1}, $w/c = 0.35$.\label{table:sims}}
% Following tex.stackexchange 419647
% Note also tex.stackexchange 207457
\begin{tabular}{ |c|c|c|c|c| } 
 \hline
 \multirow{2}{*}{Simulation(s)} & \multirow{2}{*}{AR} & \multirow{2}{*}{$t/c$} &  \multirow{2}{6em}{Number~of triangles} & \multirow{2}{*}{$\lambda$} \\
  & & & & \\ 
 \hline
 Fig.~\ref{fig:experimpics} & NA & 0.0087 & 23798 & 0.9, 0.85, 0.8 \\
 \hline 
  Fig.~\ref{fig:theory_fig} & NA & 0.0087 & 23798 & 0.8 \\
 \hline 
 Fig.~\ref{fig:simulation1},~\ref{fig:energy_fig} & NA &  0.00017 & 943178 & 0.8 \\
 \hline
 \multirow{5}{*}{Fig.~~\ref{fig:simulation2}} & 1.00 & 0.0087  & 58696 & 0.8\\
 \cline{2-5}
  & 1.75 & 0.0087 & 33495 & 0.8\\
 \cline{2-5}
  & 2.50 & 0.0087 & 23287 & 0.8\\
 \cline{2-5}
  & 3.25 & 0.0087  & 17864 & 0.8\\
 \cline{2-5}
  & 4.00 & 0.0087  & 14355 & 0.8 \\
 \hline
\end{tabular}
\end{table}

\begin{figure}[h]
	\centering
	\includegraphics[width=0.8 \columnwidth ]{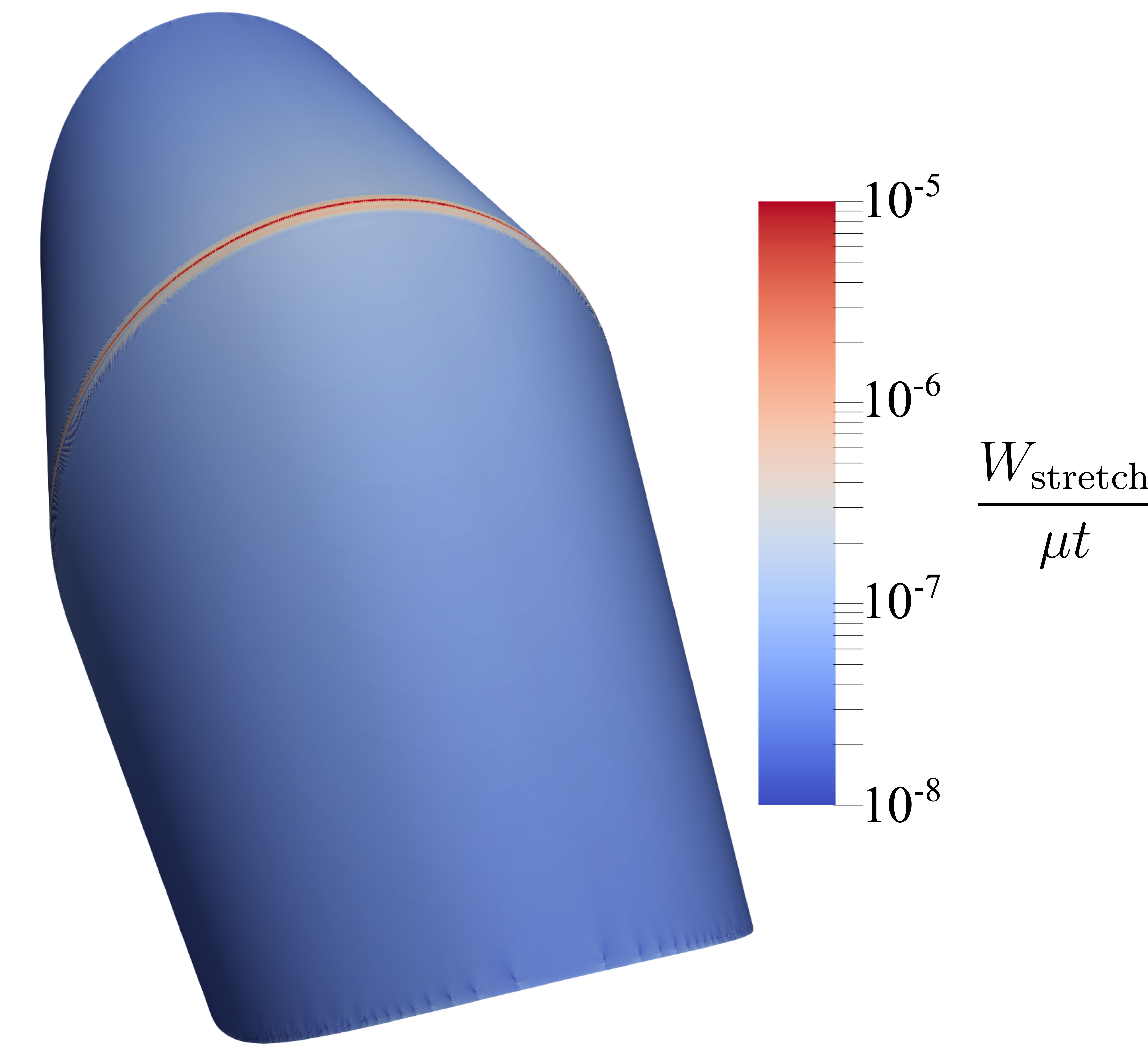}
	\caption{Simulated $\lambda=0.8$ cone complement cutout from Fig.~\ref{fig:simulation1}, here coloured by non-dimensionalized stretch energy density, quantifying the degree of non-isometry with respect to $\bar{a}$. The extremely low values far from the ridge line confirm that the surface is very nearly an isometry of $\bar{a}$ there, while the values near the ridge are much higher due to the non-isometric blunting effect.}\label{fig:energy_fig}
\end{figure}

\section{Gauss-Bonnet for finite fused cones}\label{sec_appendix_GB}
\begin{figure}[]
	\centering
	\includegraphics[width=\columnwidth]{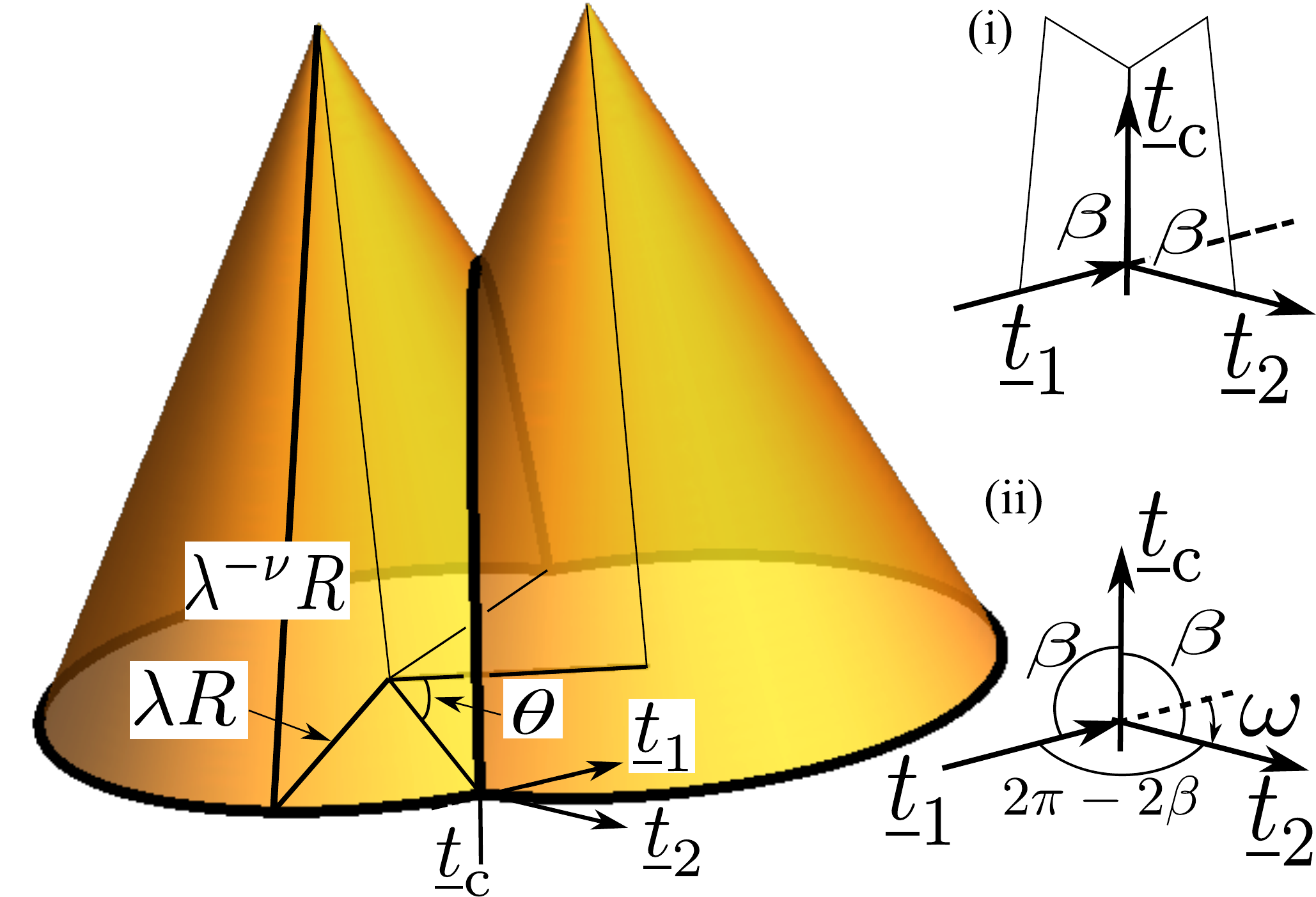}
	\caption{Two cones with the construction to evaluate the Gauss-Bonnet result for the integrated crease GC. Insets: (i) the tangent planes to the two cone flanks where they meet at the crease as it intersects with the basal perimeter. The tangent planes are flattened to a common plane by rotation about $\vec{t}_c$ as an axis. (ii) Flattened into a common plane, the angle between $\vec{t}_1$ and $\vec{t}_2$ is $\omega$.}
	\label{fig:GBcone}
\end{figure}

We sketch the use of the Gauss-Bonnet theorem to evaluate the integrated crease GC. The twin cones have the topology of a disc, see Fig.~\ref{fig:twocones}(a), and hence an Euler characteristic $\chi = 1$. The GB theorem relates the integrated GC, the boundary integral of the geodesic curvature and the sum of the tangential jump angles along the boundary, all in the target space:
\be \label{eq:GB-applied}
%\label{eq:GB}
\begin{split}
2\pi &= \int \d A_A K_A + \int \d  s_A k_g + \Sigma_i \omega_i\\
 &=   \Omega + 2\times \left[  2\pi(1-\lambda^{1+\nu})  + \right.  \\ & \;\;\;\;\ \left. + (2\pi - 2 \theta) \lambda R /(\lambda^{-\nu} R)  + \omega \right].
 \end{split}
 %\label{eq:GB-applied}.
\ee
See Fig.~\ref{fig:GBcone}.
The integrated GC is that of the two tips and that, $\Omega$, in the crease (for which we solve). The geodesic curvature of the basal boundary is $1/(\lambda^{-\nu} R)$, which can be seen by slitting the cone along a generator, and flattening it. A radial, in-material line from the tip to the base is shown, marked with its length $\lambda^{-\nu}R$,  the inverse length of which gives the geodesic curvature. The length of the basal contour is $2 \times (2\pi - 2 \theta) \lambda R$, where $\lambda R$ is the in-space radius that sweeps through an angle $(2\pi - 2 \theta)$ in the base of each cone. The angle $\theta$ depends on the length of the crease through the value of the parameter $u$ at its end as $\cos\theta = 1/\cosh(u)$; see the right triangle marked with $\theta$ in Fig.~\ref{fig:twocones}(a). The jump through angle $\omega$ from the tangent $\vec{t}_1$ of the basal boundary from the left cone to that $\vec{t}_2$ of the right cone occurs twice, once at each end of the crease where it meets the base. The tangent $\vec{t}_1$ meets that of the crease $\vec{t}_c$ at that point at an angle $\beta$ shown in insets to Fig.~\ref{fig:GBcone} given by $\cos\beta = \lambda^{1+\nu}/\sqrt{\lambda^{2(1+\nu)} +\sinh^2(u)}$. Inset (ii) also shows the two tangent planes of the fusing cones at this point flattened to a single plane by rotating about the crease tangent there that is common to them both, the angle $\beta$ of course not changing under that operation. The three tangent vectors now being in a plane, one can see that the required jump angle is $\omega = \pi - (2\pi - 2\beta) =2 \beta -\pi $. Solving eqn.~(\ref{eq:GB-applied}) for $\Omega$ yields eqn~(\ref{eq:creaseGC}). 

 Alternatively, one can obtain the geodesic curvature on each side of the crease (either by explicit calculation in the target space, or by considering splay-bend in the reference state director field \cite{defective_nematogenesis}) and then use GB locally (by flattening the tangent planes as we have done here) to directly obtain a once-integrated GC density along the crease.
% \begin{table}
% \end{tabular}
% \end{table}

\bibliography{references.bib}

\end{document}